# Statistical detection of Josephson, Andreev, and single quasiparticle currents in scanning tunneling microscopy

Wonhee Ko[1], Eugene F. Dumitrescu[2], Petro Maksymovych[1*]

[1]*Center for Nanophase Materials Sciences, Oak Ridge National Laboratory, Oak Ridge, Tennessee 37831, USA*

[2]*Computational Sciences and Engineering Division, Oak Ridge National Laboratory, Oak Ridge, Tennessee 37831, USA*

[*]*Email: maksymovychp@ornl.gov*

Abstract

We present a method to identify distinct tunneling modes in tunable superconducting tunnel junction composed of superconducting tip and sample in scanning tunneling microscope. Combining the measurement of the relative decay constant of tunneling current extracted from *I-V-z* spectroscopy with its statistical analysis over the atomic disorders in the sample surface, we identified the crossover of tunneling modes between single quasiparticle tunneling, multiple Andreev reflection, and Josephson tunneling with respect to the bias voltage. The method enables one to determine the particular tunneling regime independently of the spectral shapes, and to reveal the intrinsic modulation of Andreev reflection and Josephson current that would be crucial for quantum device application of superconductors.

The tunnel junction between superconductors is the heart of modern quantum information devices. Superconductor qubits, which are one of the most promising scalable qubits right now, heavily rely on the Josephson effect, which is quantum tunneling of Cooper pairs between the two superconductors [1,2]. The superconducting tunnel junction can also probe the nature of superconductivity, because tunneling behavior is highly dependent on the type of superconductors, such as either s-wave or d-wave [3,4], and trivial or topological [5].

Inside the superconducting tunnel junction, the interplay between electrons and Cooper pairs results in various modes of tunneling, i.e., single quasiparticle tunneling, multiple Andreev reflection (MAR) and Josephson tunneling [6]. One important factor that defines the mode of tunneling is the tunneling barrier, like its height [7,8] and capacitance [9,10]. However, in superconducting tunnel devices, tunneling barrier is not controllable, and so the systematic study of the crossover between these tunneling modes for the change in barrier properties is difficult. Scanning tunneling microscopy (STM), in contrast, can precisely control the tunneling distance down to picometer resolution by controlling its tip height, and determine the geometry of the tunneling junction in atomic scale by imaging the surface of superconductor. Thus, STM provides a fitting platform to study the behavior of various tunneling modes with respect to the tunneling barrier [11-16].

In this letter, we used STM with a superconducting tip to make a superconducting tunnel junction with vacuum tunneling barrier that is controllable to picometer precision. The spectroscopy based on *I-V-z* characteristics allows us to extract the relative decaying constant of the tunneling current with respect to the tip height and display the crossover between the single quasiparticle tunneling, MAR, and Josephson tunneling. Moreover, the relative decay constant of different tunneling modes displays distinct spatial distribution relative to the atomic disorders, whose statistical analysis gives complementary information for the transitions between the tunneling modes. Theoretical calculation with tight-binding simulation shows the broad variability of tunneling probability of Andreev reflection as a function of quasiparticle energy, whose resonant tunneling at the superconducting gap edge ultimately translates into intermediate decay rate-constant between single quasiparticle and Josephson tunneling. Our results indicate that the superconducting tunnel junction can be viewed as a parallel resistor circuit of various tunneling regimes, which can be turned on/off or amplified by the junction geometry and tunneling energy.

The experiment was performed in SPECS JT-STM operated at UHV condition ($< 10^{-10}$ mbar) and base operating temperature of 1.2 K. The surface of Pb(110) single crystal was cleaned by repeating sputtering-annealing cycles several times. The condition for sputtering was ion energy of 1 kV, Ne gas pressure of about $1 \times 10^{-5}$ mbar, emission current of 10 mA, ion current of 12 μA, and duration of 15 min, and the annealing condition was 150 °C for 40 min. The STM tip was coated with Pb by applying bias voltage of 100 V to the tungsten tip and then dip it into the Pb(110) single crystal while limiting the maximum current to be 100 μA with a 10 MΩ resistor in series [17]. Conversely, Pb on the tip can be removed by field emission on Au(111) with bias voltage of 100 V. The *I-V* curves were acquired by sweeping DC bias without any additional electrical signals. d*I*/d*V* curves were obtained by numerical differentiating the *I-V* curves and Gaussian smoothing.

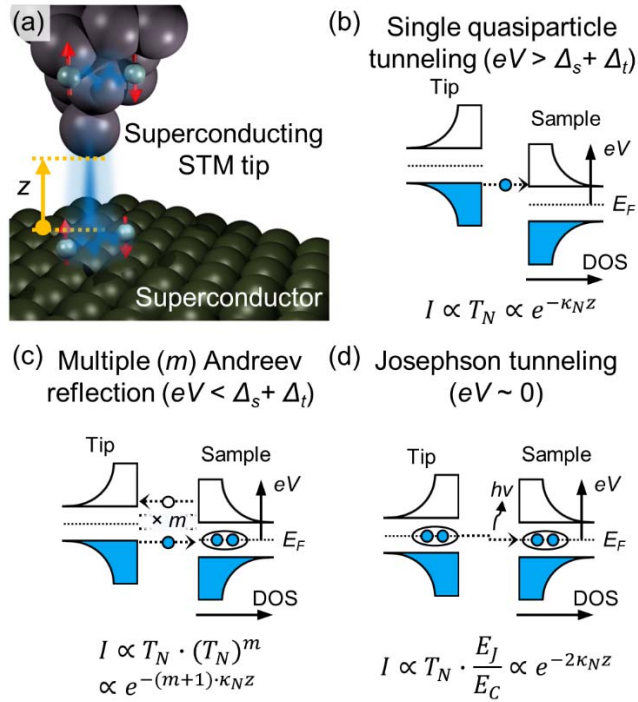

FIG 1. (a) Schematic of tunable superconductor tunnel junction composed of a superconducting STM tip and a superconducting sample. (b)-(d) Band diagrams of single quasiparticle tunneling, multiple Andreev reflection, and Josephson tunneling, respectively. Under each diagram, the expected formula of exponentially decaying current is written for the asymptotic case of large $z$.

Figure 1(a) shows the schematic of the tunable superconducting tunnel junction realized by STM. STM tip height $z$ is controllable in a picometer precision, which enables precise measurement of the decay of tunneling current $I$ with respect to $z$. Theoretical models based on Bardeen's formalism [18] predicted that single electron tunneling probability $T_N$ decays exponentially with $z$ in the asymptotic case of large $z$ [19]. In case of superconducting tunnel junction, single quasiparticle tunneling dominates when the bias is larger than the sum of the sample superconducting gap $\Delta_s$ and the tip superconducting gap $\Delta_t$ [Fig. 1(b)] [11,20]. The decay constant of single quasiparticle tunneling is same as the one in normal state [18], which we denote as $\kappa_N$. However, when the bias is lower than $\Delta_s + \Delta_t$, single quasiparticle tunneling is suppressed due to vanishing density of states. The tunneling current now involves Cooper pairs, leading to MAR and Josephson tunneling. The decay constants of these mechanisms are generally different from single quasiparticle tunneling (as schematically shown in Fig.1). Cuevas et al. showed that the tunneling probability of MAR process with $m$ times reflection is proportional to $(T_N)^{m+1}$ by using Keldysh nonequilibrium Green function method, and so the decay constant of MAR becomes $(m+1)\cdot\kappa_N$ [Fig, 1(c)] [21,22]. Meanwhile, Josephson tunneling in STM happens in the dynamical Coulomb regime, where a small capacitance of STM tunnel junction forces Cooper pairs to tunnel sequentially while emitting microwave photons [Fig. 1(d)] [9,14,15]. In this regime, inelastic tunneling process adds another factor of $E_J/E_C$ to the tunneling probability, where $E_J$ is Josephson energy and $E_C$ is capacitive charging energy [23]. The Ambegaokar–Baratoff formula shows that $E_J$ is proportional to the tunneling probability $T_N$ [7], while $E_C$ stays almost constant with sub-nanometer shift of $z$ because it mostly depends on the macroscopic tip geometry [14,24], so the decay constant becomes $2\kappa_N$.

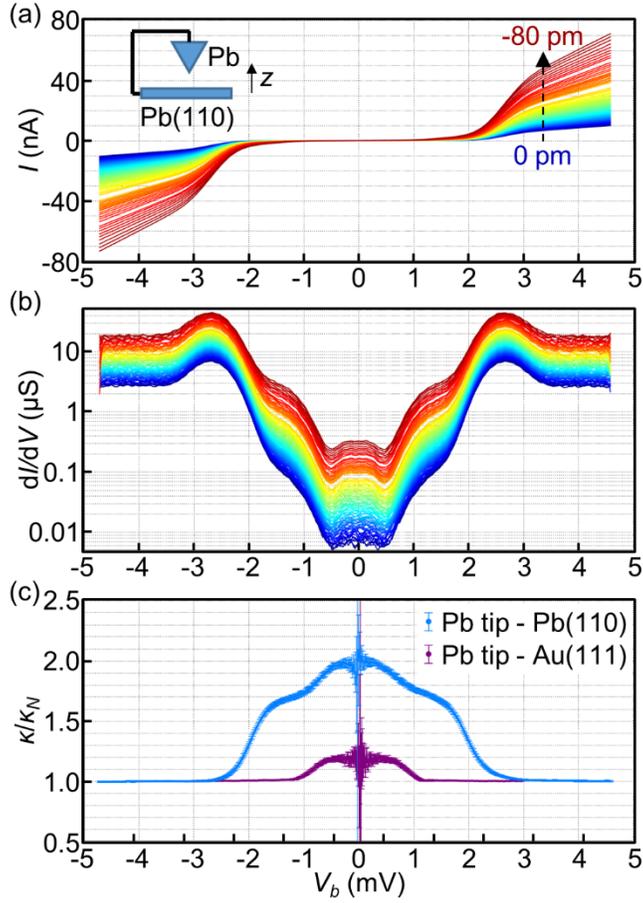

FIG 2. (a) *I-V* curves between Pb tip and Pb(110) for every 1 pm step as the tip approaches from 0 to -80 pm. $z = 0$ is defined as the tip height at $V_b = 4.6$ mV and $I = 10$ nA. (b) d*I*/d*V* curves obtained by numerical differentiation of *I-V* curves in (a). (c) Relative decay constant $\kappa/\kappa_N$ for Pb tip on Pb(110) and Au(111).

Figure 2 shows the *I-V-z* spectroscopy measurement on the Pb tip – Pb(110) junction [Inset of Fig. 2(a)]. Figure 2(a) displays the *I-V* curves while approaching the tip to the sample surface. d*I*/d*V* spectra from the numerical derivatives of *I-V* curves show clear superconducting gap between the coherence peaks at ±2.6 mV [Fig. 2(b)], as expected from the $\Delta_s \approx 1.35$ meV in Pb and $\Delta_t$ slightly less than $\Delta_s$ [11,15]. Inside the superconducting gap, the structure of d$I$/d$V$ first rapidly decreases with approaching the Fermi level, but also reveals an increase of tunneling conductance at the Fermi level particularly for spectra acquired with closest proximity between the tip and the surface. The bias lower than the superconducting gap indicates that the origin of the intra-gap features is MAR or Josephson tunneling, or perhaps both. We note that the feature at the Fermi level is broader compared to the previous reports of Josephson currents in STM junction [11,16], possibly due to the effective junction temperature of 3.2 K that is slightly higher than the base temperature (as estimated by fitting the superconducting gap of Pb(110) [25]).

The decay constant $\kappa(V_b)$ was extracted by fitting $I$-$z$ or d$I$/d$V$-$z$ curves to the exponential function [25]. We further defined the relative decay constant as $\kappa/\kappa_N$, where $\kappa_N \equiv \kappa(V_{max})$. In Fig. 2(c), we plot $\kappa/\kappa_N$ extracted from $I$-$z$ curves of Pb tip – Pb(110) junction. It is clearly shown that $\kappa/\kappa_N = 1$ for $e|V_b| > \Delta_s + \Delta_t$ as expected for single quasiparticle tunneling, but as $e|V_b|$ becomes smaller than $\Delta_s + \Delta_t$, it increases from 1 to almost exactly 2 around $V_b = 0$ mV, as expected for Josephson tunneling [14,23]. The transition between $\kappa/\kappa_N = 1$ and $\kappa/\kappa_N = 2$ proceeds through two nearly flat plateaus at around $|V_b| = 1.0 \sim 1.5$ mV with $\kappa/\kappa_N = 1.7 \pm 0.05$. The spike feature right at 0 mV is an artifact due to very small currents and the resulting uncertainty in the fitting of the $I$-$z$ curves.

The value of $\kappa/\kappa_N$ larger than 1 demonstrates that the tunneling process involves Cooper pairs, like the case of MAR and Josephson tunneling. The expected biases for single Andreev reflection and Josephson tunneling are $e|V_b| \sim \Delta_s \approx 1.35$ meV and $e|V_b| \sim 0$ meV, respectively [15], so we can tentatively assign the plateaus in $\kappa/\kappa_N$ accordingly. However, $\kappa/\kappa_N = 1.7$ for single Andreev reflection deviates from the expected value of 2 [21,22], and so we need further confirmation that this value is indeed from Andreev reflection and explore the origin of the deviation. In a control measurement of $\kappa/\kappa_N$ with Pb tip – Au(111) junction, $\kappa/\kappa_N = 1$ outside the superconducting gap but rose to 1.2 inside the gap. This observation qualitatively supports our assignment of $1 < \kappa/\kappa_N < 2$ for single Andreev reflection, and that the double step feature in Pb tip – Pb(110) junction is not an artifact of the Pb coated tip.

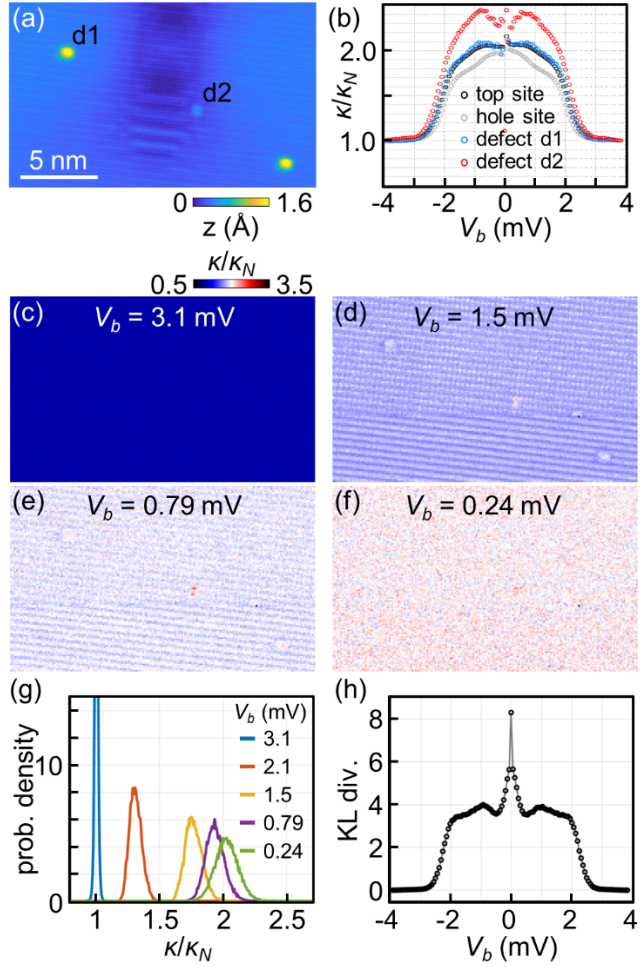

FIG. 3. (a) STM topograph of Pb(110) ($V_b = 4.6$ mV, $I = 10$ nA). (b) $\kappa/\kappa_N$ curves of four different types of sites in (a). (c)-(f) The maps of the relative decay constant at the biases labeled on top. (g) The probability density from the histogram of the $\kappa/\kappa_N$ values in the map at certain bias. (h) KL divergence between the probability density of $\kappa/\kappa_N$ values at maximum bias and other biases, centered on the mean value.

To gain complementary insight into the crossover of tunneling mechanisms for different $\kappa/\kappa_N$ regimes, we observe the distinct behavior of Andreev reflection and Josephson tunneling with respect to disorder. It is known that MAR strongly depends on the detailed atomic structure of the junction [8,26], while Josephson tunneling does not because Cooper pairs are delocalized over the coherence length (~ 80 nm in Pb) [27]. Therefore, they should be discernable via spatial distribution in heterogeneous samples.

To probe the feasibility of differentiating the mechanisms via disorder, we acquired the distance dependent spectroscopy in a wide area of the Pb(110) surface with natural population of defects [Fig. 3(a)]. The STM topograph displays two types of surface defects, labeled d1 and d2, and one subsurface defect that generates scattering pattern around the center. Figure 3(b) shows four representative $\kappa/\kappa_N$ curves: taken on top of the surface Pb atoms (top site), on the hole between the surface Pb atoms (hole site), and on defects d1 and d2. For $|V_b| = 0.5 \sim 2$ mV where MAR dominates, there is large variation of $\kappa/\kappa_N$. Even on the flat surface, single Andreev reflection at $|V_b| = 1.3$ meV exhibits atomic-scale variation where the $\kappa/\kappa_N$ of the hole site and top site varies from 1.7 to 1.9. Defects display more vivid contrast, as demonstrated in the $\kappa/\kappa_N$ of d2 which rises to 2.5 at $|V_b|= 0.7$ mV. $\kappa/\kappa_N > 2$ indicates that the atomic structure of the defect facilitates second Andreev reflection. Such enhancement of higher order MAR highly depends on the atomic structure of the defects, as demonstrated in the $\kappa/\kappa_N$ of defect d1 whose value is lower than 2 for all bias. However, as $V_b$ approaches 0 mV, $\kappa/\kappa_N$ in all positions converges to 2, whose spatial uniformity supports the transition from MAR to Josephson tunneling.

For statistical analysis of the effect of disorder, we then acquired *I-V-z* curves on the regularly spaced grid. The maps of $\kappa/\kappa_N$ at specific bias in Fig. 3(c-f) and their corresponding histograms in Fig. 3g confirm the observations from the individual spectra: for single quasiparticle tunneling, the map is very uniform with the values near 1 [Fig. 3(c)]; in the regime of MAR, $\kappa/\kappa_N$ displays atomic scale variation that follows the lattice of Pb(110) and atomic defects [Fig. 3(d),(e)], giving rise to broad distribution with multiple peaks and long tail in the histogram; and at near 0 mV the values of $\kappa/\kappa_N$ become more uniform [Fig. 3(f)], albeit with still relatively broad distribution (see supplemental movie in [25] for the full dataset of $\kappa/\kappa_N$ maps and histograms). Note that there was a tip change at about 2/5 point in y axis of the $\kappa/\kappa_N$ maps, but the qualitative behavior of $\kappa/\kappa_N$ that we described above does not change.

The crossover of mechanisms can be effectively captured by calculating the Kullback-Leibler (KL) divergence between the probability density of distributions of the $\kappa/\kappa_N$ values at varying bias and the maximum bias, such as the ones in Fig. (g) from the numerical approximation of $\kappa/\kappa_N$ histograms, after centering them at the mean value [Fig. 3(h)] [28]. The remarkably sharp transitions as a function of bias are evident. Most notably, the transition between MAR and Josephson regime occurs at $|V_b| \sim 0.5$ mV in the KL divergence as a steep rise, which is also consistent with the bias where $\kappa/\kappa_N$ maps become uniform and loses atomic features [25]. We note that in the calculation of the KL divergence we intentionally subtracted the mean value of the distributions, so as to focus the comparison on the variation rather than averaged value of the $\kappa/\kappa_N$. Thereby, the statistical analysis provides independent confirmation of our hypothesis of the distinct behavior of MAR and Josephson current over disorder and displays the transition in $\kappa/\kappa_N$ distribution more clearly.

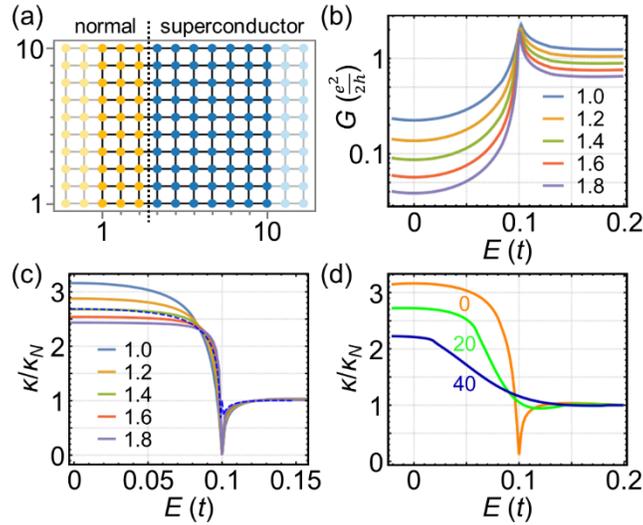

FIG. 4. (a) Schematic of the lattice structure of normal-superconductor junction for transport conductance simulation. (b) The conductance versus energy for different barrier heights. The superconducting gap was chosen as $0.1t$, where $t$ is the hopping integral. (c) The relative decay constant for different barrier heights (shown in legends, in units of $t$). (d) The relative decay constant graphs with increasing thermal broadening (legend shows the width of the Gaussian kernel used to broaden the conductance curve in units of $0.002t$).

To simulate Andreev reflection as a function of basic properties of a tunnel junction, we utilized Kwant code for tight-binding transport calculations [29]. Following a basic algorithm, the Bogoliubov-de Gennes particle-hole symmetric Hamiltonian was implemented in a spinless system without magnetic field on a square lattice. The system is schematically shown in Fig. 4(a). The scattering region is localized between columns 3 and 4, with the superconducting gap being non-zero in columns 4-10 (blue dots), and the normal lead in columns 1-3 (yellow dots). The shaded dots on each side denote the beginning of infinite leads attached to each region, with the same corresponding Hamiltonians as in the regions. The conductance $G$ calculated for such geometry clearly exhibits a superconducting gap and resonant Andreev reflection on the gap-edge [Fig. 4(b)]. Subsequently, we have systematically varied the properties of the tunneling barrier to calculate conductance as a function of barrier properties. The analysis of the apparent decay constants was then carried out similarly to the experiment.

As seen in Fig. 4(c), the $\kappa/\kappa_N$ exhibits a broad range, from near 0 to about 3, across the superconducting gap. The values near 0 originate from resonant Andreev tunneling at the gap-edge, where the transmission probability is enhanced as is common for generic resonant tunneling. The effect of the resonance decays as the energy decreases toward the middle of the gap. Eventually $\kappa/\kappa_N$ exceeds 2 and reaches values as high as 3. The results for Andreev reflection are similar for the cases where we tune the barrier height and barrier width, although the resonance effect is most pronounced in the barrier-height dependent calculation. It is also important to consider the effect of broadening on these results, which were calculated at 0K. A simplest broadening of ~ 2kT applied to the calculated $\kappa/\kappa_N$ reveals that the resonance-mediated decrease below 1 is quickly smeared [Fig. 4(d)]. However, the effect of the resonance is still present, producing intermediate $\kappa/\kappa_N$ between 1 and 2, over the width of the about half-the gap. Overall, these results are qualitatively similar to the experimental observations in Fig. 2(c).

Both larger and smaller than 2 values for $\kappa/\kappa_N$ in the Andreev regime allows us to propose a simple conceptual picture of the crossover of tunneling mechanisms as three parallel channels whose resistance depends on tip height $z$ and the tunneling energy $e|V_b|$. When $e|V_b| > \Delta_s + \Delta_t$ and the density of states is considerable, single quasiparticle tunneling is dominant even though the MAR and Josephson tunneling is possible. Meanwhile, in the region of the superconducting gaps, the preponderance of Josephson vs Andreev tunneling will be determined by their decay constant, with the smaller decay constant determining the dominant current. Just below the gap, the resonance effect favors single Andreev reflection. Toward zero bias, higher order MARs appear successively, but also Josephson tunneling starts to compete with MAR, and eventually crossover to Josephson tunneling happens because MAR has higher decay constant than Josephson tunneling. The crossover will be sensitive to many specifics of the tunneling junction and the measurement conditions. For example, the role of prefactors in exponential decay of tunneling current remains to be understood in future analysis. Conversely, detailed understanding of the crossover is likely to provide a new window into the properties of impurities in superconductors.

In summary, we combined the *I-V-z* spectroscopy of the STM-based superconducting tunnel junction with the statistical analysis over atomic disorders to provide the clear illustration of crossovers between single quasiparticle tunneling, MAR, and Josephson tunneling. The analysis enabled one to determine the particular tunneling regime independently of the spectral shapes, which provides a valuable complement to other methods of analysis of superconducting junctions. In particular, rigorously comparing and correlating the effects of disorder in different regimes will deepen our understanding on the nature of superconductivity. For example, revealing intrinsic modulations of the MAR and Josephson currents would provide a pathway to identify pairing symmetry [12], inhomogeneity of superfluid density [16,30,31], and possible existence of exotic quasiparticles [32,33], at the heart of modern quest for quantum and topological computing.

**Acknowledgments**

We gratefully acknowledge Stephen Jesse for support and discussion of this work. This research was conducted at the Center for Nanophase Materials Sciences, which is a DOE Office of Science User Facility. A portion of this research (WK,PM) was sponsored by the Laboratory Directed Research and

**Competing Interests statement**

The authors declare no competing interests.

**References**

[1] M. Kjaergaard, M. E. Schwartz, J. Braumüller, P. Krantz, J. I. J. Wang, S. Gustavsson, and W. D. Oliver, *Superconducting Qubits: Current State of Play*, Annual Review of Condensed Matter Physics **11**, 369 (2020).
[2] P. Krantz, M. Kjaergaard, F. Yan, T. P. Orlando, S. Gustavsson, and W. D. Oliver, *A quantum engineer's guide to superconducting qubits*, Applied Physics Reviews **6**, 021318 (2019).
[3] T. Bauch, T. Lindström, F. Tafuri, G. Rotoli, P. Delsing, T. Claeson, and F. Lombardi, *Quantum dynamics of a d-wave Josephson junction*, Science **311**, 57 (2006).
[4] J. F. Liu, Y. Xu, and J. Wang, *Identifying the chiral d-wave superconductivity by Josephson †$_0$-states*, Scientific Reports **7**, 43899 (2017).
[5] Y. Tanaka, T. Yokoyama, and N. Nagaosa, *Manipulation of the majorana fermion, andreev reflection, and josephson current on topological insulators*, Physical Review Letters **103**, 107002 (2009).
[6] M. Tinkham, *Introduction to Superconductivity* (Dover Publications, 2004), Second edition edn.
[7] V. Ambegaokar and A. Baratoff, *Tunneling between superconductors*, Physical Review Letters **10**, 486 (1963).
[8] G. E. Blonder, M. Tinkham, and T. M. Klapwijk, *Transition from metallic to tunneling regimes in superconducting microconstrictions: Excess current, charge imbalance, and supercurrent conversion*, Physical Review B **25**, 4515 (1982).
[9] M. H. Devoret, D. Esteve, H. Grabert, G. L. Ingold, H. Pothier, and C. Urbina, *Effect of the electromagnetic environment on the Coulomb blockade in ultrasmall tunnel junctions*, Physical Review Letters **64**, 1824 (1990).
[10] D. V. Averin, Y. V. Nazarov, and A. A. Odintsov, *Incoherent tunneling of the cooper pairs and magnetic flux quanta in ultrasmall Josephson junctions*, Physica B: Condensed Matter **165-166**, 945 (1990).
[11] O. Naaman, W. Teizer, and R. C. Dynes, *Fluctuation dominated josephson tunneling with a scanning tunneling microscope*, Physical Review Letters **87** (2001).
[12] A. V. Balatsky, J. Šmakov, and I. Martin, *Josephson scanning tunneling microscopy*, Physical Review B - Condensed Matter and Materials Physics **64** (2001).

[13]  M. Ternes, W. D. Schneider, J. C. Cuevas, C. P. Lutz, C. F. Hirjibehedin, and A. J. Heinrich, *Subgap structure in asymmetric superconducting tunnel junctions*, Physical Review B - Condensed Matter and Materials Physics **74**, 132501 (2006).

[14]  B. Jäck, M. Eltschka, M. Assig, M. Etzkorn, C. R. Ast, and K. Kern, *Critical Josephson current in the dynamical Coulomb blockade regime*, Physical Review B **93**, 020504 (2016).

[15]  M. T. Randeria, B. E. Feldman, I. K. Drozdov, and A. Yazdani, *Scanning Josephson spectroscopy on the atomic scale*, Physical Review B **93**, 161115 (2016).

[16]  D. Cho, K. M. Bastiaans, D. Chatzopoulos, G. D. Gu, and M. P. Allan, *A strongly inhomogeneous superfluid in an iron-based superconductor*, Nature **571**, 541 (2019).

[17]  K. J. Franke, G. Schulze, and J. I. Pascual, *Competition of superconducting phenomena and Kondo screening at the nanoscale*, Science **332**, 940 (2011).

[18]  J. Bardeen, *Tunnelling from a many-particle point of view*, Physical Review Letters **6**, 57 (1961).

[19]  J. Tersoff and D. R. Hamann, *Theory and application for the scanning tunneling microscope*, Physical Review Letters **50**, 1998 (1983).

[20]  R. C. Dynes, V. Narayanamurti, and J. P. Garno, *Direct measurement of quasiparticle-lifetime broadening in a strong-coupled superconductor*, Physical Review Letters **41**, 1509 (1978).

[21]  J. Cuevas, A. Martín-Rodero, and A. L. Yeyati, *Hamiltonian approach to the transport properties of superconducting quantum point contacts*, Physical Review B - Condensed Matter and Materials Physics **54**, 7366 (1996).

[22]  J. C. Cuevas and W. Belzig, *Full counting statistics of multiple andreev reflections*, Physical Review Letters **91** (2003).

[23]  G. L. Ingold, H. Grabert, and U. Eberhardt, *Cooper-pair current through ultrasmall Josephson junctions*, Physical Review B **50**, 395 (1994).

[24]  C. R. Ast, B. Jäck, J. Senkpiel, M. Eltschka, M. Etzkorn, J. Ankerhold, and K. Kern, *Sensing the quantum limit in scanning tunnelling spectroscopy*, Nature Communications **7**, 13009 (2016).

[25]  See Supplemental Material at http://link.aps.org/... for further analysis details and full dataset of the relative dacay constant maps.

[26]  J. Brand, P. Ribeiro, N. Néel, S. Kirchner, and J. Kröger, *Impact of Atomic-Scale Contact Geometry on Andreev Reflection*, Physical Review Letters **118**, 107001 (2017).

[27]  K. M. Bastiaans, D. Cho, D. Chatzopoulos, M. Leeuwenhoek, C. Koks, and M. P. Allan, *Imaging doubled shot noise in a Josephson scanning tunneling microscope*, Physical Review B **100**, 104506 (2019).

[28]  S. Kullback and R. A. Leibler, *On Information and Sufficiency*, Ann. Math. Statist. **22**, 79 (1951).

[29]  C. W. Groth, M. Wimmer, A. R. Akhmerov, and X. Waintal, *Kwant: A software package for quantum transport*, New Journal of Physics **16**, 063065 (2014).

[30]  Z. Du, H. Li, S. H. Joo, E. P. Donoway, J. Lee, J. C. S. Davis, G. Gu, P. D. Johnson, and K. Fujita, *Imaging the energy gap modulations of the cuprate pair-density-wave state*, Nature **580**, 65 (2020).

[31]  M. H. Hamidian et al., *Detection of a Cooper-pair density wave in $Bi_2Sr_2CaCu_2O_{8+x}$*, Nature **532**, 343 (2016).

[32]  K. T. Law, P. A. Lee, and T. K. Ng, *Majorana Fermion Induced Resonant Andreev Reflection*, Physical Review Letters **103**, 237001 (2009).

[33]  S. Zhu et al., *Nearly quantized conductance plateau of vortex zero mode in an iron-based superconductor*, Science **367**, 189 (2020).

*Supplemental Material of*

# Statistical detection of Josephson, Andreev, and single quasiparticle currents in scanning tunneling microscopy

Wonhee Ko[1], Eugene F. Dumitrescu[2], Petro Maksymovych[1*]

[1]*Center for Nanophase Materials Sciences, Oak Ridge National Laboratory, Oak Ridge, Tennessee 37831, USA*

[2]*Computational Sciences and Engineering Division, Oak Ridge National Laboratory, Oak Ridge, Tennessee 37831, USA*

[*]*Email: maksymovychp@ornl.gov*

## 1. Effective electron temperature of the STM tunnel junction from the gap fitting

To estimate the effective temperature of the STM tunnel junction, we measured superconducting gap of Pb(110) with normal metal tip and fit the curve with the Dynes formula:

$$\rho_{BCS}(E) = \begin{cases} 0 \quad (if\ |E| < \Delta), \\ \mathrm{Re}\left[\dfrac{E - i\Gamma}{\{(E - i\Gamma)^2 - \Delta^2\}^{1/2}}\right] \ (if\ |E| > \Delta) \end{cases}$$

$$\rho(E, T_{eff}) = \int \rho_{BCS}(E) f'(E, T_{eff}) dE$$

, where $\rho$ is electron density of states, $\Delta$ is superconducting gap, $\Gamma$ is intrinsic broadening, $T_{eff}$ is effective electron temperature, and $f(E,T)$ is Fermi-Dirac distribution given as

$$f(E, T) = \frac{1}{e^{E/k_B T} - 1}$$

Figure S1 shows the result of fitting d$I$/d$V$ spectrum to $\rho(eV, T_{eff})$, which gives $T_{eff}$ = 3.2 K, $\Delta$ = 1.5 meV, and $\Gamma$ = 4.7 × 10$^{-7}$ meV.

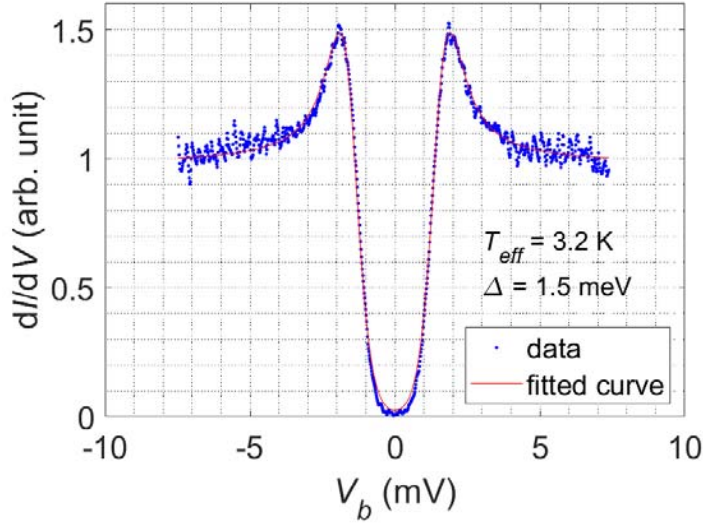

FIG. S1. d$I$/d$V$ spectra of Pb(110) measured by normal metal tip (blue dots), and fitting of the curve with the Dynes formula (red line).

## 2. Extraction of the decay constants from *I-z* fitting and their error analysis

To extract the decay constant $\kappa$ from the *I-V-z* curves, we rearranged the data to plot *I-z* or d$I$/d$V$-*z* curves at different biases and fitted the curves to exponential curves. Fig. S2(a) shows the examples of the *I-z* curves and overlayed exponential fitting, and Fig. S2(b) shows $\kappa/\kappa_N$ extracted from the fitting and $R^2$ of the fit. $R^2$ is almost 1 for all bias ranges except the very near zero where the current reached the noise limit and has large fluctuation [see lowest curve in Fig. S2(a)]. The $R^2$ analysis shows the fitting is nearly perfect for most bias ranges except the very near zero, which also explains the large error bar in $\kappa/\kappa_N$ at around the zero bias. Figure S2(c-d) shows the same fitting process to d$I$/d$V$-*z* curves and $\kappa/\kappa_N$ extracted from the fit. Here, there is no divergence of fit around 0 mV because of non-zero d$I$/d$V$ at zero bias. However, overall $R^2$ is smaller for d$I$/d$V$-*z* fitting than the *I-z* fitting due to the better signal-to-noise ratio of *I*.

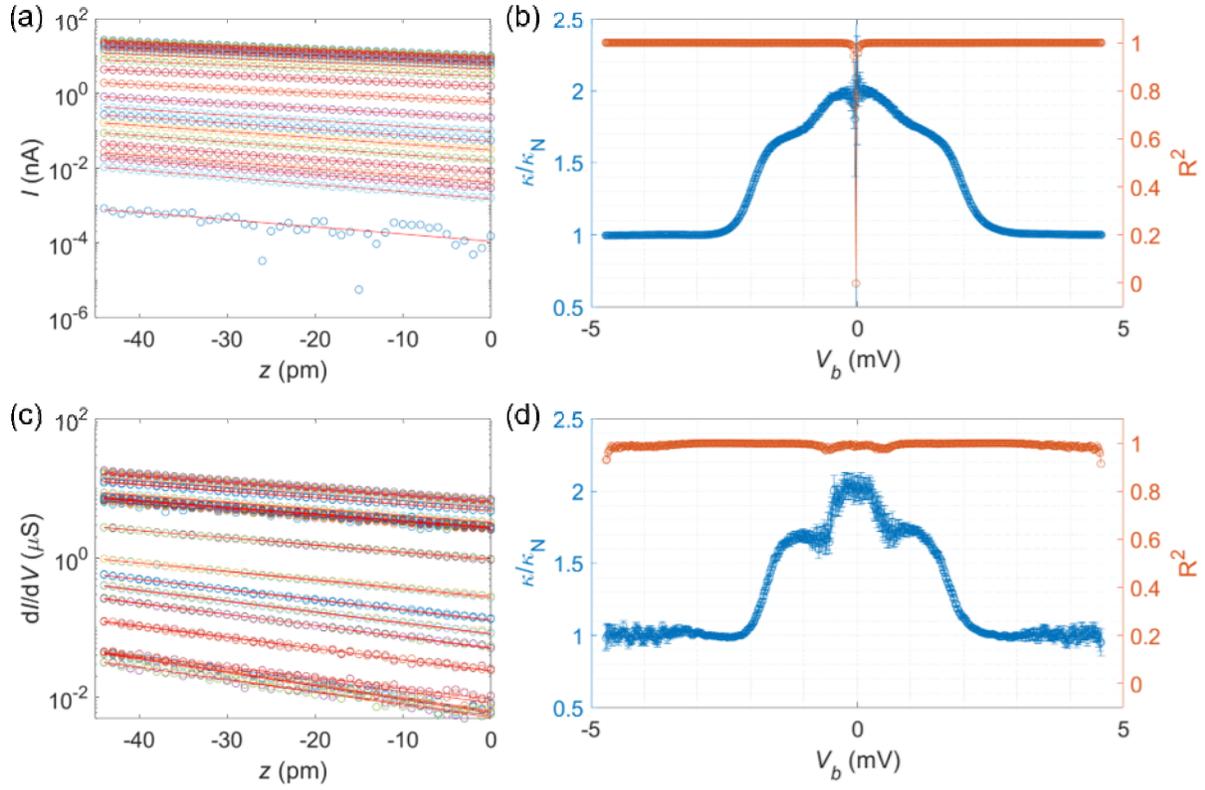

FIG. S2. (a) *I-z* curves for different $V_b$ fitted to the exponential function (red lines). Note that *y*-axis is in log scale. (b) The relative decay constant $\kappa/\kappa_N$ and the $R^2$ from the fitting in (a). (c) d*I*/d*V*-*z* curves for different $V_b$ fitted to the exponential function (red lines). Note that *y*-axis is in log scale. (d) The relative decay constant $\kappa/\kappa_N$ and the $R^2$ from the fitting in (c).

## 3. Maps and histograms of $\kappa/\kappa_N$ for full bias range

Movie S1 displays all $\kappa/\kappa_N$ maps for full bias range (left) and the probability density of $\kappa/\kappa_N$ from their histograms (right).